\documentclass[aps,english,showpacs,twocolumn,pra,superscriptaddress]{revtex4-1}
\usepackage{amsfonts}
\usepackage{amssymb}
\usepackage{amsmath}
\usepackage{graphicx}
\usepackage{epsfig}
\usepackage{color}
\usepackage{float}
\usepackage{subfigure}

\begin{document}

\title{Berry curvature inside parity-time-symmetry protected exceptional
surface}
\author{P. Wang}
\affiliation{Department of Mathematics and Physics, North China Electric Power
University, Beijing 102206, China}
\author{L. Jin}
\email{jinliang@nankai.edu.cn}
\author{Z. Song}
\email{songtc@nankai.edu.cn}

\affiliation{School of Physics, Nankai University, Tianjin 300071, China}
\begin{abstract}
A three-dimensional non-Hermitian Hamiltonian with parity-time
symmetry can exhibit a closed exceptional surface (EP surface) in momentum
space, which is a non-Hermitian deformation of the degeneracy line (DL).
Since the degeneracy line lacks an internal space, the distributions of
Berry curvature inside the EP surface becomes particularly intriguing. This
paper studies the distributions taking a torus-like EP surface as an
example. In a meridian cross-section, the Berry connection exhibits a
vortex-like field with only angular components, while the Berry curvature is
perpendicular to this cross-section; in a equatorial cross-section, the
Berry curvature forms a closed curve surrounding the central genus. Both
Berry connection and curvature converge along the coplanar axis and diverge
at the surface. We find the Berry flux depends on the radius of the
integration region and is not quantized inside the EP torus. Approaching the
surface, the Berry flux tends to infinity and the dynamical phase oscillates
violently. We point that the streamlines of Berry curvature can be used to
estimate the zero or non-zero Berry flux. We generalize the above patterns
to the case of EP surfaces with complex shapes, and present a proposal of
realizing the EP surface in an electrical circuit. Our research outcomes
enhance the comprehension of EP surfaces and the topological characteristics
of non-Hermitian systems with parity-time ($\mathcal{PT}$) symmetry.
\end{abstract}

\maketitle


\noaffiliation

\section{Introduction}

The complex band structure and non-orthogonal eigenmodes induced by
non-Hermiticity exhibit numerous intriguing properties \cite%
{BenderRPP,Moiseyev,LFeng,LonghiEL,EINP,Miri,LYangNM,UedaAdv2020,KunstRMP2021,GMaNRP2022}%
. The non-Hermitian phase transition occurs at the exceptional point (EP)
where two or more eigenstates coalesce. The EP is unique and causes many
exotic phenoma, such as polynomially increasing power \cite%
{WPPRA,Ge18,LJinPRL18,XHS2} and sensitive dynamics near the EPs \cite%
{LJin09,LJin10,CTChan1,LJinPRA18,CTChan3,Ganainy18,SMZhang,JinPRB20}.
Considerable attention has been focused on the non-Hermitian topological
phase. In Hermitian systems, the appearance of edge states \cite{XLC}
depends on the topological properties of the bulk system, known as the
bulk-boundary correspondence (BBC) \cite{HCWuPRL24}. In non-Hermitian
systems, the BBC may be invalided by the non-Hermiticity associated with the
non-Hermitian skin effect \cite%
{WangPRL2018,ZWangPRL086803,TLeePRL2016,JLPRB2019,Longhi066602,Sato1,ZW1,
CFang,BYan,LLWan,SPKouPRB165420,PXuePRL230402,CFangPRL226402}%
. Exotic edge modes localized on the single boundary and the topological
number from a non-Block bulk predict the topological phase transitions of
the corresponding non-Hermitian systems \cite%
{WHCPRB2019,ZW2,Slager,CFangPRL,ZYang}. Methods for characterizing the
topology of non-Hermitian bands are investigated \cite%
{Ueda1,LFu,Duan,DL1,SM,Nori,KLZhang,YKe}. A visualization of the
topological properties is proposed \cite{JBG2,XMYang,HCWu}. The origin and
properties of non-Hermitian edge modes are further studied \cite{DL3,YZhang}
and the symmetry and classification of topological phases are reestablished
\cite{WQC,SHF1,JYL,Sato3,Sato4,Ueda2,Yoshita3,CHLee06974,HParkPRB2022,SPKouPRB165139}. The non-Hermitian
extension of Hermitian models exhibits alternative topological phases \cite%
{CWZ1,CWZ3,SL6,CLee06323,DL2,SChen05688,Longhi064303,LCXiePRB21,SPKouPRB104306}, including
the non-Hermitian Aubry-Andre-Harper models \cite{ZLi,YXu020201}, the
Su-Schrieffer-Heeger models \cite%
{NCandemir,SChen,CY4,WPSR2017,LJinPRA2017,HCWuPRB2021}, and the
non-Hermitian disordered topological systems \cite%
{SLZ,SL3,CY3,CZhang10652,XCXie}. The Floquet topological phase \cite%
{Levitov,LZhou18,Joglekar18} and quantum walks \cite{XDZhang,SL1,Fritzsche}
are extended to non-Hermitian systems. The interplay between time-periodic
driving fields and the presence of gain, loss, or nonreciprocal effects can
lead to the emergence of topological phases exclusive to non-Hermitian
Floquet systems \cite{LWZ1,LWZ2,LBF,JBGong045415,SChen2022}. The
deformation of the contour specific to a topological invariant is
demonstrated to accommodate the non-Hermiticity of the underlying
non-interacting Hamiltonian in question \cite{QHLeePRB2019}. In addition,
many studies have focused on the novel topological nature induced by
non-Hermiticity \cite%
{CWZ2,Ezawa1,Nori2,JHou13184,Yoshida1,Franz,DHX,SR,BJYang,RLu,SChen00554,JHB,GCLR,CY1,CY2,YZhai,PXuePRL230401,PXueNC2293,SPKouPRR033196}%
.

The high dimensional EP structure is a noteworthy problem. Exceptional rings
(EP rings) have been intensively discussed theoretically and experimentally
\cite{RY,BZhen,SHF2,ZXZEL,YChongNPhoton,Hatsugai,PXuePRL026404}. An EP ring can be
analogous to a vortex filament, and the curl field related to the vortex
filament is equivalent to the Berry connection \cite{ZXZEL}. In
three-dimensional (3D) momentum space, non-Hermitian Hamiltonians with
combined parity and time reversal symmetry spontaneously meet conditions for
the appearance of exceptional surfaces (EP surfaces) \cite{Yoko,OL}. The EP
surface is stable as long as the protecting symmetry is preserved \cite%
{BergholtzPRB}. The EP surface inherits the topological properties of the
degenerate line (DL); the nodal volume, which represents bulk Fermi arcs in
3D space, indicates the remarkable control of the density of states (DOS)
\cite{OL}. The topological properties of the EP surface can also be
characterized by $Z_{2}$ topological invariants, and a stable zero-gap
quasi-particle state is protected by symmetry and topology \cite{Yoko}. In a
high-dimensional parameter space, a hypersurface where the system remains at
an EP improves the robustness and enhances the sensitivity of EPs, and a
non-Hermitian sensor can be designed on the basis of the hypersurface \cite%
{GanainyPRL}. A non-Hermitian Bardeen-Cooper-Schrieffer (BCS) Hamiltonian
with a weak complex interaction possesses an EP surface in the quasi-partile
Hamiltonian, and non-Hermiticity induces the breaking down of superfluidity
and exhibition of reentrant behavior \cite{Ueda3}. The EP surface affects
magnetic responses in a Hubbard model; the sharp local density of states
(LDOS) at the Fermi energy for sublattices with weak correlations results in
the local magnetic susceptibility of strong sublattice dependence \cite%
{Kimura}. Experimentally, the EP surface can be observed on a magnon
polariton platform, and the EP surface can be conveniently tuned to coalesce
into an anisotropic exceptional saddle point \cite{DJinA}.

Motivated by recent theoretical advances in non-Hermitian topological
systems, we investigate the distribution of Berry curvature inside the EP
surface. The Berry curvature is gauge-invariant and related to the
topological properties of EP surfaces. In this paper, we investigate a
two-band non-Hermitian system with parity-time ($\mathcal{PT}$) symmetry and
a closed EP surface in 3D momentum space. The general expression of the
Berry curvature defined under the biorthogonal basis reveals that the EP
surface separates the zero and non-zero Berry curvature. A Hamiltonian with
a torus-like EP surface is exemplified. The topological properties of the EP
surface are encoded in the distributions of the Berry curvature in the
meridians and equatorial cross-sections. In the meridians cross-section, the
Berry connection acts as a planar vortex field and the direction of the
Berry curvature is perpendicular to this cross-section; in the equatorial
cross-section, Berry curvatures form closed curves. Both Berry connection
and curvature are convergent at the coplanar axis and divergent at the EP
surface. The surface integral of the Berry curvature yields a non-quantized
Berry flux. The numerical simulation implies that the non-quantized Berry
flux is consistent with the dynamics phase accumulated in the adiabatic
evolution, and both of them oscillates violently near the EP surface. The
Berry flux can be evaluated by the distribution of Berry curvature. Berry
flux is nonzero if the Berry curvatures have the same direction in the
meridians cross-section otherwise vanishes. These patterns can be
generalized to the EP surfaces with complicated geometries. Finally, rather
than the realization in coupled resonators \cite{XHS}, we point that the EP
surface can be measured in a electrical circuit.

The remainder of the paper is organized as follows: In Sec. \ref%
{Non-Hermitian two-band system}, we introduce the 3D\ {$\mathcal{PT}$%
-symmetric }non-Hermitian two-band system. In Sec. \ref{Berry connection and
Berry curvature}, we present the formal expression of Berry connection and
curvature. In Sec. \ref{Torus-like EP surface}, we introduce a concrete
model to exhibit the Berry curvature inside a torus-like EP surface. The
non-quantized Berry flux is elucidated from the distribution of the Berry
curvature. In Sec. \ref{Adiabatic evolution}, the adiabatic evolution is
implemented. In Sec. \ref{EP surface with complicated geometry}, a
topological system that possesses more complicated EP surface is further
discussed. In Sec. \ref{Experimental scheme in electrical circuit}, the
proposal of realizing the EP surface is given in the electrical circuits. In
Sec. \ref{Conclusion}, we summarize the results.

\section{Non-Hermitian two-band system}

\label{Non-Hermitian two-band system}We consider a non-Hermitian two-band
Hamiltonian in the momentum space $\mathbf{k=}\left\{
k_{x},k_{y},k_{z}\right\} $,%
\begin{equation}
h_{\mathbf{k}}=\mathbf{B}\left( \mathbf{k}\right) \cdot \mathbf{\sigma },
\label{Hamiltonian}
\end{equation}%
{where $\mathbf{\sigma }=$}$\{\sigma _{x}\mathbf{,}\sigma _{y}\mathbf{,}%
\sigma _{z}${$\}$ is the Pauli matrix; the component }${\{B_{x}\left(
\mathbf{k}\right) ,B_{y}\left( \mathbf{k}\right) \}}${\ of the auxiliary
field $\mathbf{B}\left( \mathbf{k}\right) $ is the real and periodic
function of $k=\left\{ k_{x},k_{y},k_{z}\right\} $; and the other component $%
B_{z}=i\gamma $ is a constant, which is introduced as the gain and loss.}%
$h_{\mathbf{k}}$ possesses the $\mathcal{PT}$-symmetry $\left[
\mathcal{PT},h_{\mathbf{k}}\right] =0$, where $\mathcal{P}=\sigma _{x}$ is
the parity operator and $\mathcal{T}$\ is the time-reversal operator that $%
\mathcal{T}^{-1}i\mathcal{T}=-i$. The {eigenvalues of }$\mathcal{PT}${%
-symmetric systems are either real numbers or complex conjugate pairs
respectively associated with }$\mathcal{PT}$-symmetry unbroken or broken
eigenstates, respectively. Considering the specific form of the band in Eq. (%
\ref{Hamiltonian}), i.e $\pm \sqrt{B_{x}^{2}+B_{y}^{2}-\gamma ^{2}},$ the
complex conjugate pairs are reduced to purely imaginary numbers. The
eigenstate expressions involve parameters defined in terms of energy,
therefore the real/imaginary eigenvalues make the expressions more concise.
In addition, $h_{\mathbf{k}}$ is the pseudo anti-Hermitian $\sigma _{z}h_{%
\mathbf{k}}\sigma _{z}^{-1}=-h_{\mathbf{k}}^{\dagger }$~{\cite{LJinCPL21}}.
It is straightforward to check that $h_{\mathbf{k}}^{\dagger }\sigma
_{z}\left\vert \phi \right\rangle =-\varepsilon \sigma _{z}\left\vert \phi
\right\rangle ,$ where $h_{\mathbf{k}}\left\vert \phi \right\rangle
=\varepsilon \left\vert \phi \right\rangle $. This implies that $\sigma
_{z}\left\vert \phi \right\rangle $ becomes the left eigenstate
corresponding to the right eigenstate $\left\vert \phi \right\rangle $ when $%
\varepsilon $ is purely imaginary. These characteristics of $h_{\mathbf{k}}$
simplify the calculations in the following text and are reflected in Sec. %
\ref{Berry connection and Berry curvature}.

In the Hermitian case ($\gamma =0$), the band degeneracy is determined by
the following equations%
\begin{equation}
B_{x}\left( \mathbf{k}\right) =B_{y}\left( \mathbf{k}\right) =0.  \label{DP}
\end{equation}%
$B_{x}\left( \mathbf{k}\right) =0$ and $B_{y}\left( \mathbf{k}\right) =0$\
each represent a surface in the 3D momentum space. The intersection of two
surfaces is the degeneracy line (DL). The topological properties
of the DL are captured by the topological number Berry flux or winding
number. The former is the integral of the Berry connection on a closed
circle, while the latter is obtained by dividing the Berry flux by $\pi $.
The Berry flux is quantized to $\pi $ ($0$) if the closed circle is (not)
linked with the DL \cite{XDai,CKC}. In the presence of gain and loss for $%
\gamma \neq 0$, the DL becomes an EP surface. The EP surface\ is the
zero-energy surface in the form of%
\begin{equation}
\gamma ^{2}=B_{x}^{2}(\mathbf{k})+B_{y}^{2}(\mathbf{k}).  \label{EPsurface}
\end{equation}%
We consider the case in which Eq. (\ref{EPsurface}) describes a closed 2D
surface in the 3D momentum space at the selected $\{B_{x}\left( \mathbf{k}%
\right) ,B_{y}\left( \mathbf{k}\right) \}$. In this situation, the energy is
real outside the closed EP surface and is purely imaginary inside the closed
EP surface. We regard the purely imaginary region as the nodal volume
wrapped by the EP surface. These data serve as the 3D bulk Fermi arcs \cite%
{OL}. $\mathcal{PT}$-symmetry protects the EP surface\ which inherits the
Berry flux of the DL \cite{Yoko,OL}.\ In this work, we focus on the Berry
curvature distributions inside and outside the EP surface.

\section{Berry connection and Berry curvature}

\label{Berry connection and Berry curvature}
\begin{figure*}[th]
\includegraphics[bb=0 0 941 271, width=16 cm, clip]{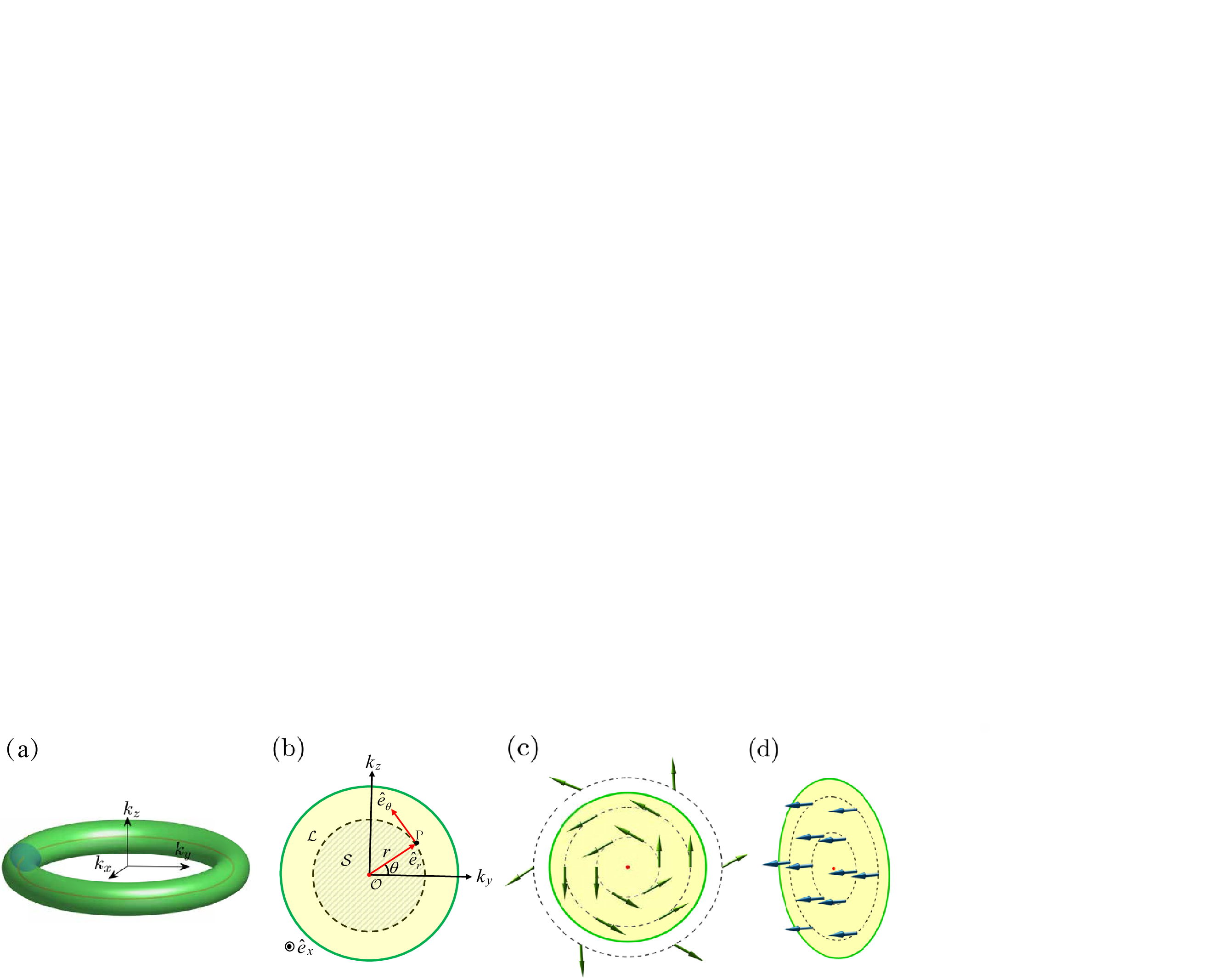}
\caption{(a) Torus EP surface (green) at $\protect\gamma =0.05$, $a\approx
3.2$, $m\approx 4.05$. The dark disk is the representative cross-section $%
S_{V}$ and the red co-planar circular axis is the DL. (b) Schematic diagram
of the the cross-section $S_{V}.$ (c) Berry connection and (d) Berry
curvature in the cross-section $S_{V}$.}
\label{fig1}
\end{figure*}

This section provides the general expressions of Berry connections and
curvatures inside and outside the EP surface.

We first calculate the eigenstates of the Bloch Hamiltonian under
biorthogonal norm. The right eigenstates $\left\vert \phi _{\pm }^{\mathrm{R}%
}\right\rangle $ of the Bloch Hamiltonian satisfy $h_{\mathbf{k}}\left\vert
\phi _{\pm }^{\mathrm{R}}\right\rangle =\varepsilon _{\pm }\left\vert \phi
_{\pm }^{\mathrm{R}}\right\rangle $ and the left eigenstates $\left\vert
\phi _{\pm }^{\mathrm{L}}\right\rangle $ satisfy $h_{\mathbf{k}}^{\dagger
}\left\vert \phi _{\pm }^{\mathrm{L}}\right\rangle =\varepsilon _{\pm
}^{\ast }\left\vert \phi _{\pm }^{\mathrm{L}}\right\rangle $. They are
normalized under the biorthogonal norm $\left\langle \phi _{\alpha }^{%
\mathrm{L}}\right. \left\vert \phi _{\alpha }^{\mathrm{R}}\right\rangle =1$ $%
(\alpha =+/-)$. The EP surface serves as a boundary separating the real and
complex energies. The Bloch Hamiltonian possesses an entirely real spectrum
outside the EP surface (i.e. the $\mathcal{PT}$-symmetry unbroken phase) and
possesses an entirely imaginary spectrum inside the EP surface (i.e. the $%
\mathcal{PT}$-symmetry broken phase). For the geometric features of the
lower band $\left\vert \phi _{-}^{\mathrm{R}}\right\rangle $ with energy $-%
\sqrt{B_{x}^{2}+B_{y}^{2}-\gamma ^{2}}$, in the unbroken $\mathcal{PT}$%
-symmetry region $\gamma ^{2}<B_{x}^{2}+B_{y}^{2}$, the right and left
eigenstates are in the form%
\begin{eqnarray}
\left\vert \phi _{-}^{\mathrm{R}}\right\rangle &=&[e^{i\left( -\alpha -\beta
\right) },1]^{\mathrm{T}}\leftrightarrow -\varepsilon ,  \label{Phi_R_out} \\
\left\vert \phi _{-}^{\mathrm{L}}\right\rangle &=&[e^{i\left( \alpha -\beta
\right) },-1]^{\mathrm{T}}/\Omega \leftrightarrow -\varepsilon ,
\label{Phi_L_out}
\end{eqnarray}%
respectively, where $\varepsilon =\sqrt{B_{x}^{2}+B_{y}^{2}-\gamma ^{2}}%
,\alpha $ and $\beta $ are determined by $\tan \alpha =\gamma /\varepsilon $
and $\tan \beta =B_{y}/B_{x}$ respectively, and $\Omega =-2ie^{i\alpha }\sin
\alpha $. In the broken region $\gamma ^{2}>B_{x}^{2}+B_{y}^{2}$, the right
and left eigenstates are in the form
\begin{eqnarray}
\left\vert \phi _{-}^{\mathrm{R}}\right\rangle &=&[\eta e^{i\left( \frac{\pi
}{2}-\beta \right) },1]^{\mathrm{T}}\leftrightarrow -i\varepsilon ,
\label{Phi_R_in} \\
\left\vert \phi _{-}^{\mathrm{L}}\right\rangle &=&[-\eta e^{i\left( \frac{%
\pi }{2}-\beta \right) },1]^{\mathrm{T}}/\Omega \leftrightarrow i\varepsilon
,  \label{Phi_L_in}
\end{eqnarray}%
respectively, where $\varepsilon =\sqrt{\gamma ^{2}-B_{x}^{2}-B_{y}^{2}}$, $%
\eta =(\gamma -\varepsilon )/\sqrt{B_{x}^{2}+B_{y}^{2}}$, and $\Omega =2\eta
\varepsilon /\sqrt{B_{x}^{2}+B_{y}^{2}}$.

The Berry connection is defined as $\mathcal{\vec{A}}=\mathrm{Re}%
(i\left\langle \phi _{-}^{\mathrm{L}}\right\vert \vec{\nabla}\left\vert \phi
_{-}^{\mathrm{R}}\right\rangle )$ and the Berry curvature is defined as $%
\mathcal{\vec{F}}=\vec{\nabla}\times \mathcal{\vec{A}}$, where $\vec{\nabla}%
=\partial _{k_{x}}\hat{e}_{x}+\partial _{k_{y}}\hat{e}_{y}+\partial _{k_{z}}%
\hat{e}_{z}$. Therefore, the formal expressions of the Berry connection and
Berry curvature differ between the unbroken and broken $\mathcal{PT}$%
-symmetric phases. The Berry connection is complex in both the broken and
unbroken regions, the imaginary part amplifies the Dirac probability of the
adiabatic evolved state, whereas the real part is related to the topological
properties of the system \cite{ZXZPRA}. Therefore, we consider only the real
part of the Berry connection in the definition. The detailed calculations
are provided in the Appendix, and the results are presented concisely as
follows.

Inside the EP surface $\mathcal{PT}$-symmetry is broken. The components of
Berry connection $\mathcal{A}_{j}$ and Berry curvature $\mathcal{F}_{j}$ read%
\begin{eqnarray}
\mathcal{A}_{j} &=&\frac{(\varepsilon -\gamma )\left( B_{x}\partial
_{j}B_{y}-B_{y}\partial _{j}B_{x}\right) }{2\varepsilon \left(
B_{x}^{2}+B_{y}^{2}\right) },  \label{AcompIn} \\
\mathcal{F}_{j} &=&\frac{\gamma (\partial _{l}B_{y}\partial
_{i}B_{x}-\partial _{i}B_{y}\partial _{l}B_{x})}{2\varepsilon ^{3}},
\label{BcurIn}
\end{eqnarray}%
where $\partial _{j}=\partial /\partial k_{j}$ ($j=x$, $y$, $z$) and $%
\varepsilon =\sqrt{\gamma ^{2}-B_{x}^{2}-B_{y}^{2}}$. $\mathcal{A}_{j}$ and $%
\mathcal{F}_{j}$ converge at the DL in the Hermitian case (i.e., $%
B_{x}=B_{y}=0$ or $\varepsilon =0$) and are infinite at the singularity $%
\varepsilon =0$ (i.e., the EP surface).

Outside the nodal volume the $\mathcal{PT}$ symmetry holds. The Berry
connection and Berry curvature become%
\begin{equation}
\mathcal{A}_{j}=\frac{\left( \gamma B_{x}-\varepsilon B_{y}\right) \partial
_{j}B_{x}+\left( \varepsilon B_{x}+\gamma B_{y}\right) \partial _{j}B_{y}}{%
2\varepsilon \left( B_{x}^{2}+B_{y}^{2}\right) }.  \label{Aout}
\end{equation}%
\begin{equation}
\mathcal{\vec{F}}=0.  \label{BcurOUT}
\end{equation}%
Equations (\ref{BcurIn}) and (\ref{BcurOUT}) imply that the EP surface acts
as the boundary between zero and non-zero Berry curvature. To extract more
explicit information on the Berry connection and curvature, we further
simplify these formulas inside a torus-like EP surface.

\section{Torus-like EP surface}

\label{Torus-like EP surface}We use a concrete model possessing an EP
surface to study the distributions of Berry connection and curvature in the
broken region, from which the inheritance of the Berry flux is well
interpreted. {The auxiliary field $\mathbf{B}\left( \mathbf{k}\right)
=\left\{ B_{x}\text{, }B_{y}\right\} $ of the} concrete Hamiltonian is in
the form%
\begin{equation}
\left\{
\begin{array}{l}
B_{x}=f\left( k_{x},k_{y}\right) -s\cos k_{z} \\
B_{y}=s\sin k_{z}%
\end{array}%
\right. ,  \label{H}
\end{equation}%
where $f\left( k_{x},k_{y}\right) =m-a\cos k_{x}-a\cos k_{y}$, and $s=$ $1$.
The physical realization of the concrete Hamiltonian is proposed \cite{OL,LL}%
.

{The general geometric property of the EP surface is determined by the
components }${\{B_{x}\left( \mathbf{k}\right) ,B_{y}\left( \mathbf{k}\right)
\}}$. Equations (\ref{DP}), (\ref{EPsurface}), and (\ref{H}) indicate there
are two identical nodal volumes located at the $k_{z}=0$ and $k_{z}=\pi $
planes, only the former is studied for convenience. With fixed parameters $%
\{m,a\}$, Eq. (\ref{EPsurface}) implies that the maximum of $B_{y}$\ is $%
B_{y}=\gamma $\ (i.e., $s\sin k_{z}=\gamma $) in the situation $B_{x}=0$;
therefore, the maximum of $k_{z}$\ on the EP surface is $k_{z_{\text{\textrm{%
max}}}}=\arcsin (\gamma /s)$,\ and the restriction $\gamma <1$ is imposed.
In fact, if $\gamma =1$, the two EP surfaces touch at $k_{z_{\text{\textrm{%
max}}}}=\arcsin \gamma $. In addition, the EP surface possesses a mirror
symmetry with respect to the $k_{z}=0$\ plane.

The system possesses a torus-like EP surface under the appropriate
parameters (see Sec. \ref{Berry connection and curvature in the meridional
cross-section}). A schematic diagram of the torus-like EP surface is shown
in Fig. \ref{fig1}(a). The the red coplanar circular axis is DL in the
Hermitian case. Two types of cross-sections are studied in this paper, i.e.,
meridinal (equatorial) cross-sections in the form of closed disks (annulus).
The meridinal (equatorial) cross-sections represent the intersection of the
nodal volume and the vertical plane passing through the origin (equatorial
plane). For convenience, a cross-section denoted $S_{V}$ ($S_{H}$) as the
intersection of the nodal volume and the $k_{y}$O$k_{z}$ ($k_{x}$O$k_{y}$)
plane is chosen as a representative meridian (equatorial).Schematic diagrams
of $S_{V}$ and $S_{H}$ are shown in Fig. \ref{fig1}(b) and Fig. \ref{fig2}
respectively. The distributions of Berry curvature and Berry connection in
the other cross-sections are similar to those in $S_{V}$ and $S_{H}$.
Therefore, the distributions of berry curvature and berry connection inside
the nodal volume can be obtained once the distribution is given in the two
representative cross-sections.

\subsection{Distribution in the meridional cross-section}

\label{Distribution in the meridional cross-section}This section discusses
the distribution of Berry curvature in the cross-section $S_{V}$ of the
concrete model in Eq. (\ref{H}).

In the cross-section $S_{V}$, polar coordinates are used to describe the
physical quantities. As shown in Fig. \ref{fig1}(b), the green EP ring
divides the plane into two parts: the Hamiltonian has an entirely imaginary
spectrum in the yellow region inside the EP ring and an entirely real
spectrum outside the EP ring. The EP ring is subcircular with radius $\gamma
$. The circular dashed line is the energy contour $\mathcal{L}$ with radius $%
r$. The red point $O(0,k_{y_{0}},0)$ is center of the $S_{V}$ and is the
degenerate point (DP) in the Hermitian case ($\gamma =0$). The arbitrary
point $P(0,k_{y},k_{z})$\ on the contour $\mathcal{L}$ can be rewritten as $%
P(0,r,\theta )$\ in the cylindrical coordinate system, where $\theta $\ is
the included angle between position vector $P$\ and coordinate axis $k_{y}$.
The three unit vectors $\{\hat{e}_{\theta },\hat{e}_{r},\hat{e}_{x}\}$ of
the cylindrical coordinate system are presented. Under the parameter
settings given in Sec. \ref{Berry connection and curvature in the meridional
cross-section}, the Hamiltonian in the cross-section $S_{V}$ is reduced to
\begin{equation}
H=\left(
\begin{array}{cc}
i\gamma & re^{i\theta } \\
re^{-i\theta } & -i\gamma%
\end{array}%
\right) .  \label{Hsv}
\end{equation}%
It is straightforward to check that the reduced Hamiltonian $H$
obeys the {$\mathcal{PT}$-symmetry}, i.e., $\mathcal{T}\sigma _{x}H(\mathcal{%
T}\sigma _{x})^{-1}=H$. For the case with complex matrices, a numerical
result can be obtained, exhibiting a deformed but similar distribution of
Berry curvature, as shown in Sec. \ref{EP surface with complicated geometry}.%

The distributions of the Berry connection and Berry curvature in $S_{V}$\
are illustrated in Fig. \ref{fig1}(c) and (d). Inside the EP ring, the
expression of the Berry connection at the position $P(0,r,\theta )$ ($%
r<\gamma $) in Eq. (\ref{AcompIn}) is reduced to%
\begin{equation}
\mathcal{\vec{A}}\approx \frac{\varepsilon -\gamma }{2\varepsilon r}\vec{e}%
_{\theta }\text{ }(0\leqslant r<\gamma ),  \label{BCpolar}
\end{equation}%
where $\varepsilon =\sqrt{\gamma ^{2}-r^{2}}.$ The expression of the Berry
connection in the above equation is equal to the expression directly
calculated from Eq. (\ref{Hsv}). Equation (\ref{BCpolar}) indicates that the
radial component $\vec{e}_{r}$\ vanishes, and the angular component $\vec{e}%
_{\theta }$\ is non-zero, so $\mathcal{\vec{A}}$ is a planar vortex field%
\textbf{.} We show the direction of $\mathcal{\vec{A}}$ by the arrows
without considering its intensity according to Eq. (\ref{BCpolar}), and each
arrow is tangent to the energy contour $\mathcal{L}$. It is not difficult to
check that

\begin{equation}
\lim_{r\rightarrow 0}\frac{\sqrt{\gamma ^{2}-r^{2}}-\gamma }{2\sqrt{\gamma
^{2}-r^{2}}r}=0,
\end{equation}%
which indicates that \textbf{\ }$\mathcal{\vec{A}}$\textbf{\ }converges at $%
r=0 $ (i.e., DP at $\gamma =0$)\textbf{. }$\mathcal{\vec{A}}$ is divergent
at $\varepsilon =0$. Equation (\ref{BcurIn}) can be reduced to%
\begin{equation}
\mathcal{\vec{F}}\approx \frac{\gamma }{2\sqrt{\gamma ^{2}-r^{2}}}\vec{e}_{x}%
\text{ }0\leqslant r<\gamma  \label{Bcurvature}
\end{equation}%
inside the EP ring. The expression for the Berry curvature in the above
equation is equal to the expression directly calculated from Eq. (\ref{Hsv}%
). Equation (\ref{Bcurvature}) indicates that only the axial component $\vec{%
e}_{x}$ is non-zero. The Berry curvature has a divergent value at $r=\gamma $
(i.e. the EP ring) and a convergent value at $r=0$. In Fig. \ref{fig1}(d),
we exhibit the direction of $\mathcal{\vec{F}}$. As we can see, all the
arrows in $S_{V} $ point in the same\ direction, which is the normal of $%
S_{V}$. As an analogy, these arrows can be regarded as magnetic field lines,
and the total magnetic flux is the number of magnetic field lines that pass
through $S_{V}$. In addition, in two adjacent meridional cross-sections,
these arrows are connected end to end, as shown in the inset of Fig. \ref%
{fig2}, where the green solid lines denote the top view of meridional
cross-sections; and the dashed red line is the DL in the Hermitian case ($%
\gamma =0$). The arrows in all the meridional cross-sections form a closed
curve (see Fig. \ref{fig2}).

The Berry flux is related to the distribution of the Berry curvature or
Berry connection and can be used to capture the topological nature of the EP
surface,
\begin{subequations}
\begin{equation}
\Phi _{\mathbf{B}}=\oint_{\mathcal{L}}\mathcal{\vec{A}}\cdot d\vec{l}%
_{k}=\iint_{\mathcal{S}}\mathcal{\vec{F}}\cdot \mathrm{d}\mathcal{\vec{S}},
\label{BerryFlux}
\end{equation}%
where $\mathcal{S}$\textbf{\ }is the integral surface, which is the shaded
region surrounded by\textbf{\ }$\mathcal{L}$ presented in Fig. \ref{fig1}%
(b). By substituting Eq. (\ref{Bcurvature}) into Eq. (\ref{BerryFlux}), we
can obtain
\end{subequations}
\begin{equation}
\Phi _{\mathbf{B}}=\pi -\frac{\gamma \pi }{\sqrt{\gamma ^{2}-r^{2}}}.
\label{LI}
\end{equation}%
$\Phi _{\mathbf{B}}$ is divergent on the EP surface ($r=\gamma $). Therefore
the geometric phase oscillates sharply when the integration path approaches
the EP surface. Due to the divergence of the Berry connection and curvature
on the exceptional point (EP) surface, the line integral of the Berry
connection will not be equal to the surface integral of the Berry curvature
when the integration path is located in the unbroken region, that is, the
Stokes theorem does not hold. In addition, the Berry flux can be regarded as
the total magnetic flux. In Fig. \ref{fig1}(d), the uniform pointing of the
arrows indicate that the same sign contributes to the Berry flux and
therefore a non-zero Berry flux.

\subsection{Distribution in the equatorial cross-section}

\label{Distribution in the equatorial cross-section}

\begin{figure}[t]
\includegraphics[ bb=392 320 1424 836, width=8 cm, clip]{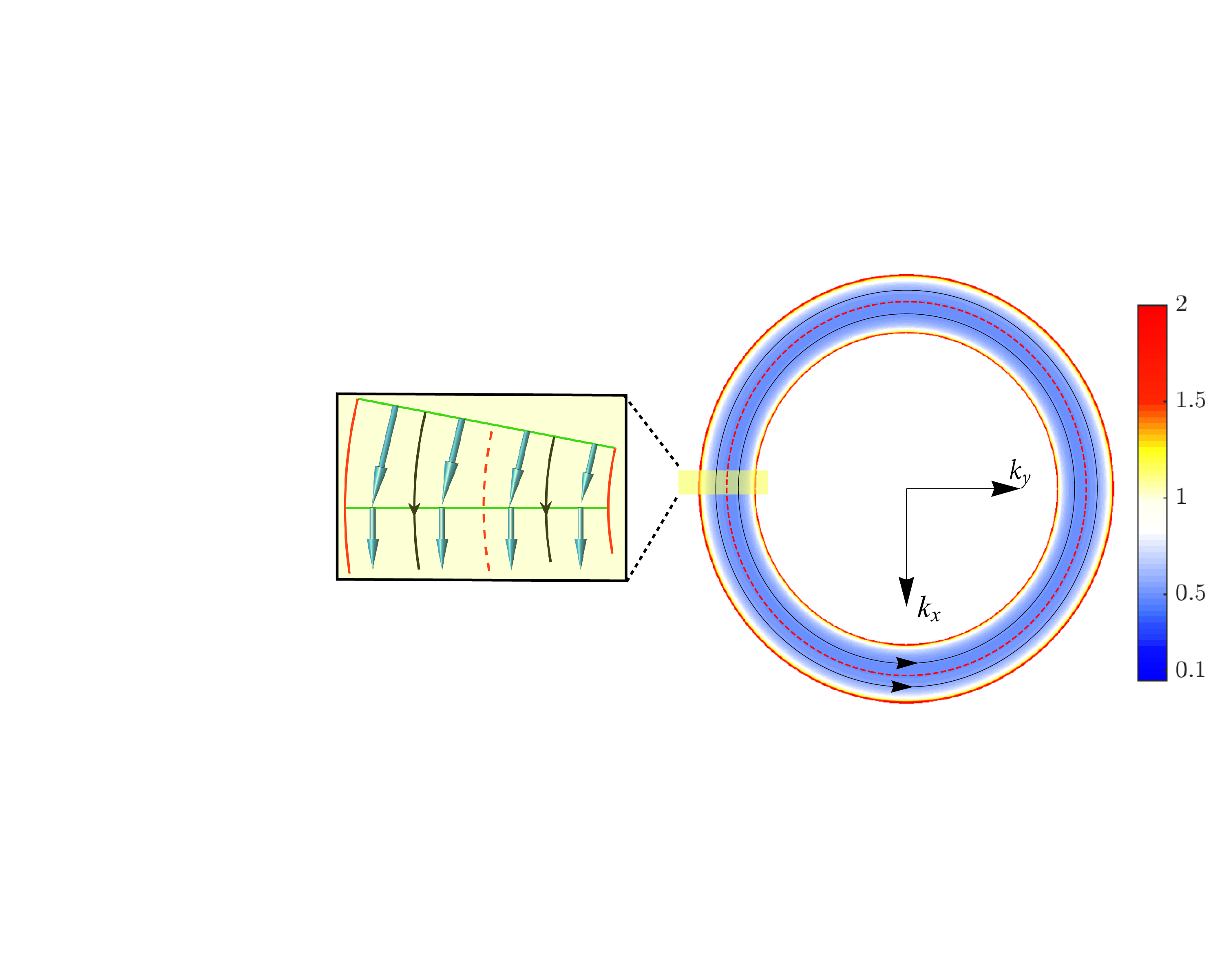}
\caption{Streamlines of Berry curvature in the equatorial cross-section.
Inset: top view of Berry curvature in $S_V$ cross-section and its adjacent
cross-section.}
\label{fig2}
\end{figure}

This section investigates the distribution of the Berry curvature in $S_{H}$.

In the equatorial cross-section $S_{H}$, the Berry curvature in Eq. (\ref%
{BcurIn}) inside the EP ring is reduced to%
\begin{equation}
\left\{
\begin{array}{l}
\mathcal{F}_{x}=s\gamma \partial _{y}B_{x}/(2\varepsilon ^{3}) \\
\mathcal{F}_{y}=-s\gamma \partial _{x}B_{x}/(2\varepsilon ^{3}) \\
\mathcal{F}_{z}=0%
\end{array}%
\right. ,  \label{BerryCurvatureII}
\end{equation}%
{Therefore, the orientation of the Berry curvature at an arbitrary point%
\textbf{\ (}$k_{x},k_{y},0$\textbf{) }in }$S_{H}${\textbf{\ }is\textbf{\ }$%
\mathcal{F}_{y}/\mathcal{F}_{x}=-(\partial _{x}B_{x})/(\partial _{y}B_{x})$.
In addition Eq. (\ref{EPsurface}) can be reduced to%
\begin{equation}
B_{x}\left( k_{x},k_{y},0\right) =\gamma ^{\prime },  \label{Curves}
\end{equation}%
in }$S_{H}$ {where\textbf{\ }$\left\vert \gamma ^{\prime }\right\vert
\leqslant \left\vert \gamma \right\vert $\textbf{. }Equation (\ref{Curves})
represents a closed curve inside the equatorial cross-section for a fixed $%
\gamma ^{\prime }$. This closed curve is the intersection between the $%
k_{z}=0$\ plane and the EP surface and is determined by replacing $\gamma $\
with $\gamma ^{\prime }$\ ($\left\vert \gamma ^{\prime }\right\vert
<\left\vert \gamma \right\vert $) in Eq. (\ref{EPsurface}). If $\gamma
^{\prime }=\gamma $, the curve is the EP ring as well as the periphery of }$%
S_{H}${.\ If $\gamma ^{\prime }$ changes from $-\gamma $\ to $\gamma $, all
the curves determined by every $\gamma ^{\prime }$\ constitute the
equatorial cross-section, and no two curves have a crossing point. The
tangent of a curve at\textbf{\ (}$k_{x}$\textbf{, }$k_{y}$\textbf{, }$0$%
\textbf{) }is $\mathrm{d}k_{y}/\mathrm{d}k_{x}=-(\partial
_{x}B_{x})/(\partial _{y}B_{x})$ as a result of complete differentiation on
both sides of Eq. (\ref{Curves})\textbf{. }Compared with the equation\textbf{%
\ }$\mathcal{F}_{y}/\mathcal{F}_{x}=-(\partial _{x}B_{x})/(\partial
_{y}B_{x})$, we conclude that the direction of\textbf{\ }Berry curvature at
the point \textbf{(}$k_{x}$\textbf{, }$k_{y}$\textbf{, }$0$\textbf{) }is
identical to the tangent of the curve passing through this point,}%
\begin{equation}
\mathcal{F}_{y}/\mathcal{F}_{x}=\mathrm{d}k_{y}/\mathrm{d}k_{x}.
\label{Orientation}
\end{equation}%
The above results hold true as long as $B_{y}$ is a function of only $k_{z}$.

The streamlines of the Berry curvature in $S_{H}$ according to Eq. (\ref%
{Curves}) are shown in Fig. \ref{fig2}. A different closed black curve is
depicted by setting different $\gamma ^{\prime }$. The red solid EP lines ($%
\gamma ^{\prime }=\gamma $) serve as the boundary separating non-zero and
zero Berry curvatures\textbf{; }the region between the two EP lines has
non-zero Berry curvature. The dashed red line ($\gamma ^{\prime }=0$)
represents the DL for the Hermitian case. The black curves ($0<\gamma
^{\prime }<\gamma $) with arrows represent the orientation of the Berry
curvature, and the background color indicates the intensity of the Berry
curvature. The intensity values are shown in the color bar. The Berry
curvature approached infinity near the EP lines. The streamlines surrounding
the hole flow anticlockwise. All the streamlines of the Berry curvature are
closed, which coincides with the equation $\nabla \cdot \mathcal{\vec{F}}=0$%
, meaning that Berry curvatures act as a field without sources. In addition,
Eq. (\ref{Orientation}) can be generalized to the other intersection between
the $k_{z}=k_{z^{\prime }}$ plane and the EP surface. The distributions
claim a clear physical correspondence for the Berry curvature and EP
surface. The Berry curvature can be analogous to magnetic lines generated by
a solenoid, and the EP surface can be connected to this solenoid. The total
magnetic flux is the number of magnetic field lines passing through certain
cross-section.
\begin{figure}[th]
\includegraphics[bb=0 0 523 250, width=8.5 cm, clip]{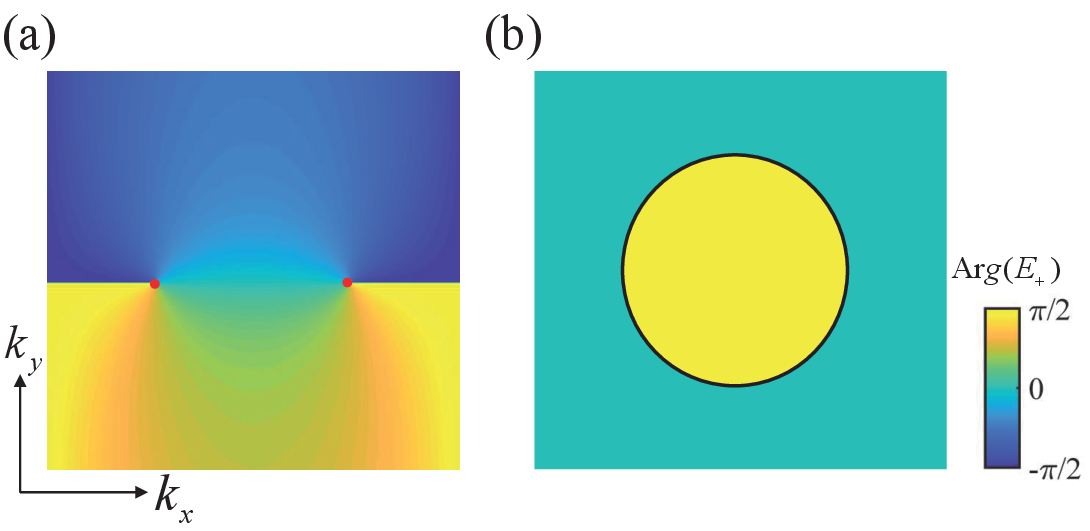}
\caption{(a) Plot of Arg($E_{+}-E_{-}$) for two isolated EPs.
The two red points represent two isolated EPs (i.e., two vortices). (b)
Plots of Arg($E_{+}-E_{-}$) for EP surface. The black line represents the EL
(i.e., a cross-section of EP surface).}
\label{fig_spectralphase}
\end{figure}

This section examined the distribution of the Berry curvature
inside the EP surface using two types of cross-sections as examples, and
calculates the Berry flux. Before moving on to the next section, there are
three points that need to be supplemented and explained:

i) The above results are obtained under the biorthogonal basis sets. Under
the Dirac basis sets, the directions of the Berry connection and Berry
curvature at any point inside the EP surface are the same but the magnitudes
are different, and the two kinds of Berry fluxes are different. Both the
Berry connection and Berry curvature under the Dirac basis sets converge on
the EP surface, therefore the Stokes theorem holds.

ii) The winding number associated with the berry connection cannot be used
to capture the topological nature of the EP surface. Figure \ref{fig1}(c)
shows the direction of $\mathcal{\vec{A}}$ denoted by arrows. The winding
numbers of the arrows along the contour $\mathcal{L}$ outside and inside the
EP ring are both non-zero. However, this non-zero winding number is not
related to the non-zero Berry flux. The Berry flux in Eq. (\ref{BerryFlux})
can be rewritten as the loop integral of the Berry connection, i.e. $\Phi _{%
\mathbf{B}}=\oint_{\mathcal{L}}\mathcal{\vec{A}}\cdot d\vec{l}_{k}$, and it
is not equal to the expression of the winding number for the Berry
connection $W=(2\pi )^{-1}\oint_{\mathcal{L}}(\mathcal{A}_{y}\nabla \mathcal{%
A}_{x}-\mathcal{A}_{x}\nabla \mathcal{A}_{y})/\left\vert \mathcal{A}%
\right\vert ^{2}d\mathbf{k}$.

iii) There is an open question that is the topological connection between
the isolated EPs and the EP surface in the context of topological number.
The nontrivial topological nature of an isolated EP depends on the scalar
field defined by the spectral phase Arg($E_{+}-E_{-}$) (see Fig. \ref%
{fig_spectralphase}(a)) \cite{LFu,QXiong}. The EP is regarded as a vortex of
the scalar field where the spectral phase cannot be effectively defined. The
topological nature of an EP can be characterized by the topological number $%
\pi $ or $1/2$, the former is the spectral phase difference accumulated when
encircling the vortex while the latter is the winding number obtained
through dividing this phase difference by $2\pi .$ Therefore EPs can be
analogous to $\pi $-vortices, which hangs together with the totpological
defect in a nematic \cite{Kamien,Slager1,Slager2,SlagerPRE2017,Slager5} or defects in TIs \cite%
{SCZhang1,SCZhang2,Slager3,SlagerPRB081117,Slager4}. The isolated EPs may merge accompanied by the
algebraic addition of topological numbers \cite{SLinPRB}. The subject of
this study is the EP surface, which is a collection of infinite EPs. Figure %
\ref{fig_spectralphase}(b) exhibits the spectral phase of EP surface. There
is no obvious evidence that the topological properties of EP surface are
related to the spectral phase. Corresponding to the same spectral pahse in
Fig. \ref{fig_spectralphase}(b), Figs. \ref{fig1}(d) and \ref{fig4}(f)
exhibit two distinct distributions of Berry curvature. The above analyses
indicate that the spectral phase cannot describe the topology of the EP
surface completely, and the winding number corresponding to EP surface is
also not equal to $\pm 1/2$. Therefore EP surface cannot be simply
understood as the merger of isolated EPs. The topological connection between
the EPs and EP surfaces deserves further investigation.

\section{Adiabatic evolution}

\label{Adiabatic evolution}

\begin{figure}[h]
\subfigure{\begin{minipage}[t]{0.5\linewidth}\includegraphics[bb=80 170 510
619, width=4.05 cm, clip]{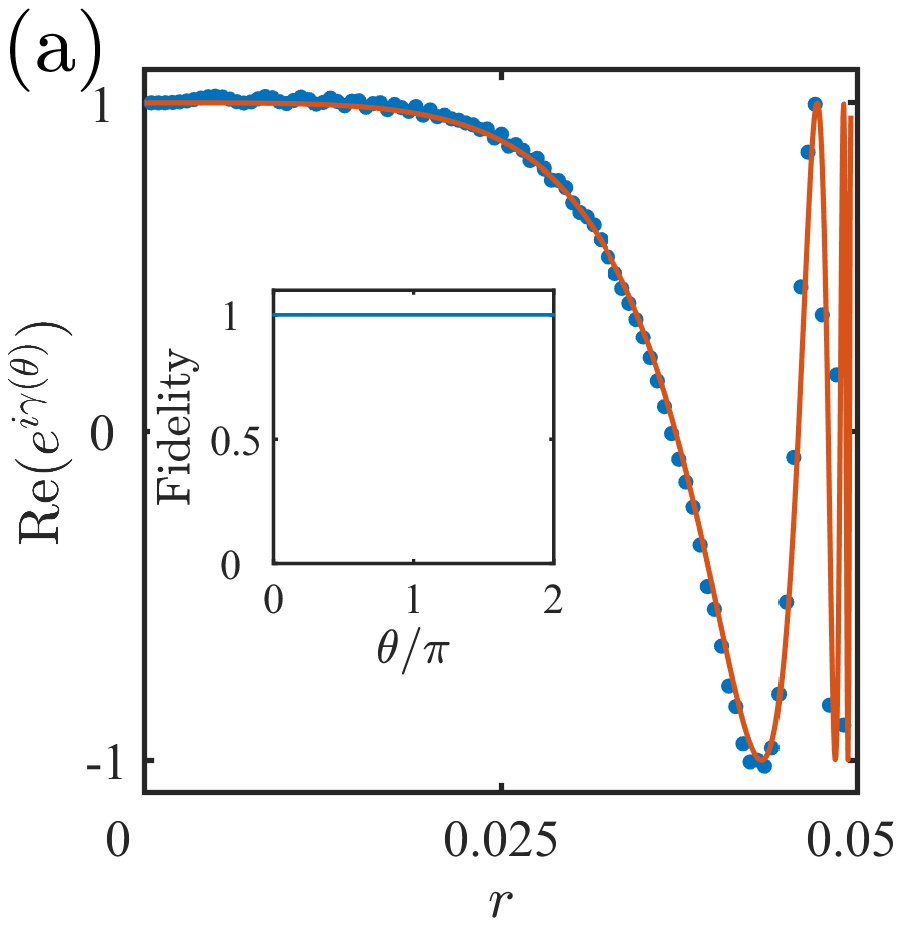} \end{minipage}}%
\subfigure{\begin{minipage}[t]{0.5\linewidth}\includegraphics[bb=0 0 425
450, width=4
cm, clip]{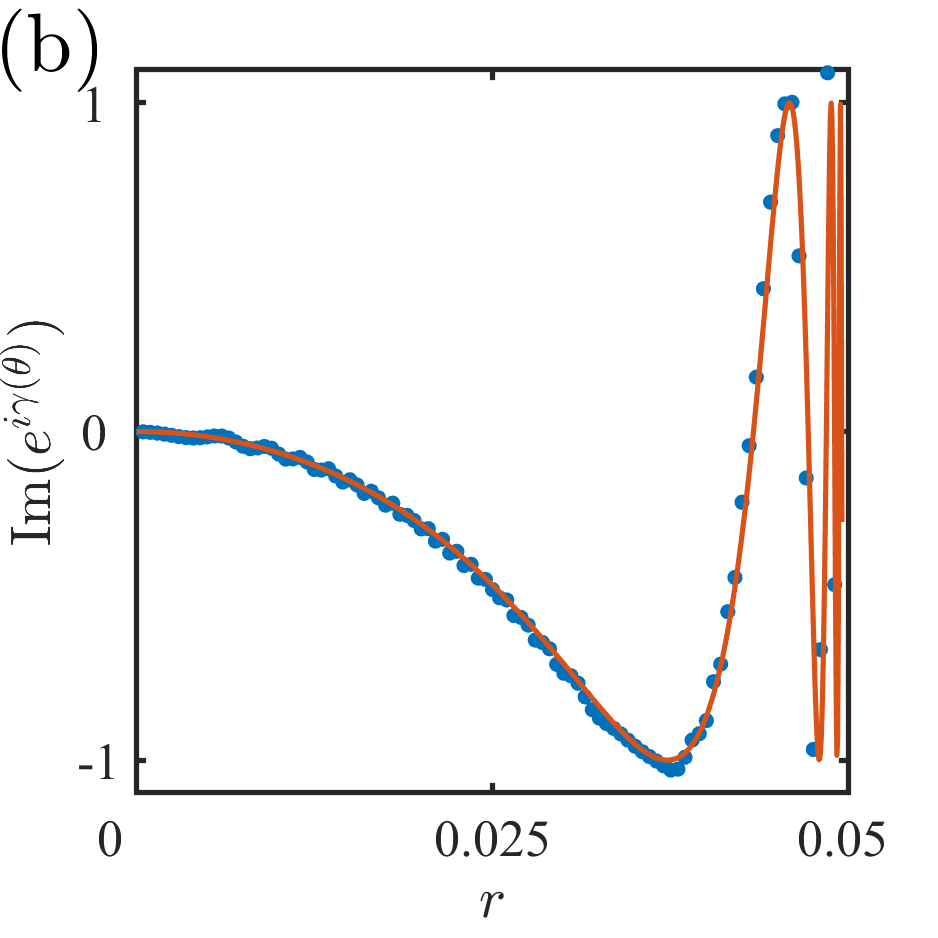}\end{minipage}}
\caption{{}Schematics of the geometric phase $e^{i\protect\gamma (\protect%
\theta )}$. (a)\ Real part and (b) imaginary part. The blue points are
numerical results and the red solid line are analytical results according to
Eq. (\protect\ref{LI}). Inset: fidelity for $r=\protect\gamma /2$. The
parameters are the same as those in Fig.\ \protect\ref{fig1}(a).}
\label{fig3}
\end{figure}

To verify the above results, we numerically simulate the adiabatic evolution
driven by the Hamiltonian in Eq. (\ref{Hsv}) and compare the geometric phase
obtained by numerical simulation and the analytical results in Eq. \ref{LI}.
\begin{figure}[th]
\includegraphics[ bb=0 0 695 995, width=8 cm, clip]{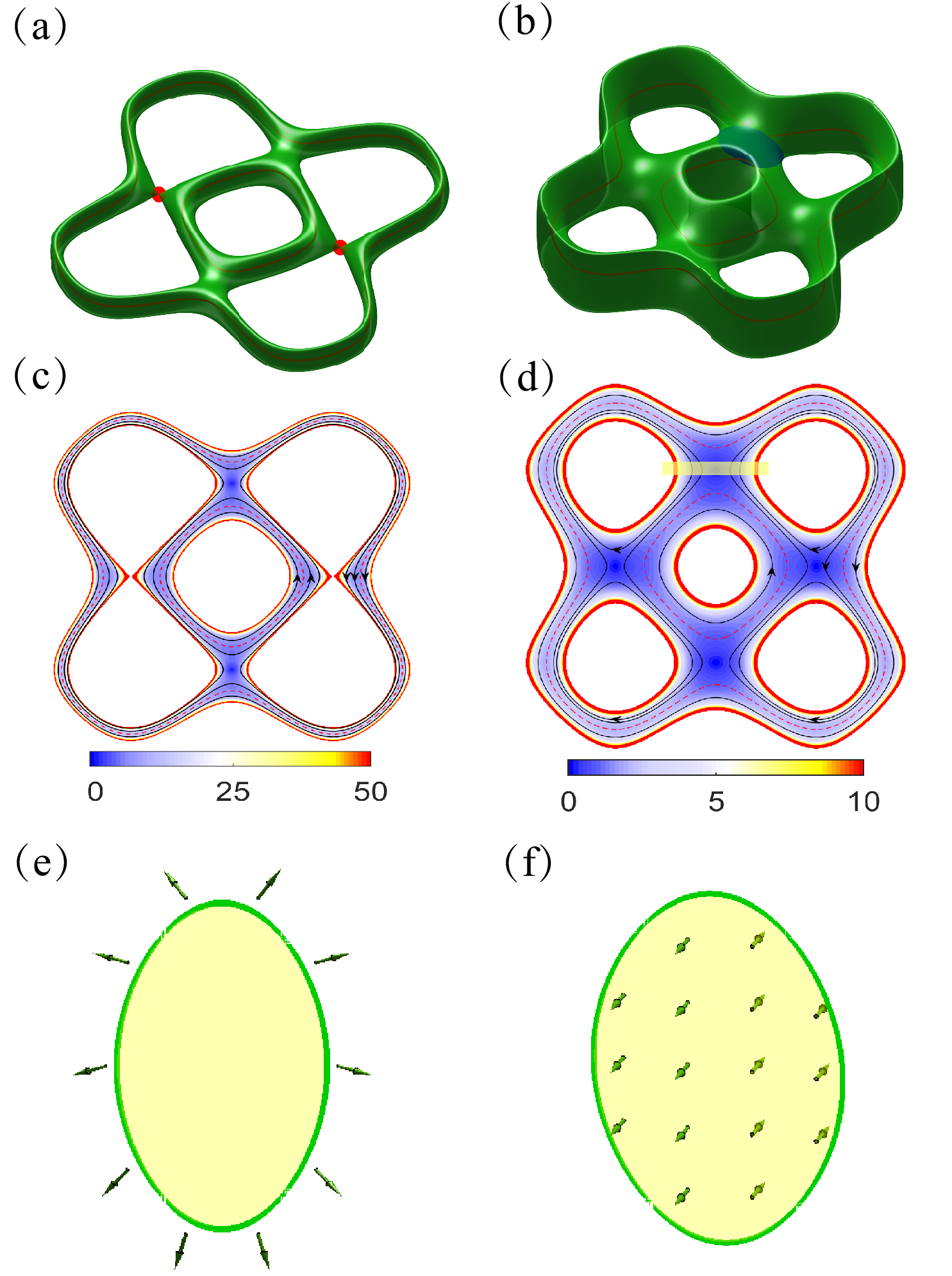}
\caption{Genus at parameters $a=2,b=2.3,c=d=2.8,m=0.3,$ (a) $\protect\gamma %
=109/140$ and (b) $\protect\gamma =1.9$. (c) and (d): Streamlines of Berry
curvature in the equatorial cross-section for the configurations in (a) and
(b), respectively. (e) and (f): Berry connection outside the EP ring and
Berry curvature inside the EP ring on a specific cross-section, this
specific cross-section is depicted in (b) and (d).}
\label{fig4}
\end{figure}
Considering the adiabatic evolution on the circular contour $\mathcal{L}$
with a radius $r$ [see Fig. \ref{fig1}(b)]. $H$ in Eq. (\ref{Hsv}) is a
periodic function of $\theta $, $H(\theta )=H(\theta +2\pi )$. The lower
band eigenstate $\left\vert \phi _{-}^{\mathrm{R}}(0)\right\rangle $ reverts
to $\left\vert \phi _{-}^{\mathrm{R}}(0)\right\rangle $ if $\theta $ varies
adiabatically from $0$ to $2\pi $, and the evolved state is the
instantaneous lower band eigenstate $\left\vert \phi _{-}^{\mathrm{R}%
}(\theta )\right\rangle $. More explicitly, the adiabatic evolution of the
initial state $\left\vert \phi _{-}^{\mathrm{R}}(0)\right\rangle $ under the
Hamiltonian $H(\theta )$ can be expressed as%
\begin{eqnarray}
\left\vert \Psi _{\lambda }^{k}(\theta )\right\rangle &=&\mathcal{T}\exp
[-i\int_{0}^{\theta }H(\theta )\mathrm{d}\theta ]\left\vert \phi
(0)\right\rangle  \notag \\
&=&e^{i\left( \alpha (\theta )+\gamma (\theta )\right) }\left\vert \phi
(0)\right\rangle ,
\end{eqnarray}%
where the dynamic phase $\alpha (\theta )$ and the adiabatic phase $\gamma
(\theta )$ have the form%
\begin{equation}
\alpha (\theta )=-\int_{0}^{\theta }\varepsilon _{k}\left( \theta \right)
\mathrm{d}\theta ,\gamma (\theta )=\int_{0}^{\theta }\mathcal{A}\left(
\theta \right) \mathrm{d}\theta .
\end{equation}%
$\mathcal{A}\left( \theta \right) $ is presented in Eq. (\ref{BCpolar}) and $%
\gamma (\theta )$ is equivalent to the Berry flux in Eq. (\ref{LI}). The
imaginary part of $\mathcal{A}\left( \theta \right) $ in Eq. (\ref{Im})
vanishes due to the invariability $\eta =(\gamma -\varepsilon )/\sqrt{\gamma
^{2}-\varepsilon ^{2}}$ on the contour $\mathcal{L}$ and therefore does not
contribute to adiabatic evolution. However $\alpha (\theta )$ is imaginary
and the Dirac probability increases exponentially. To eliminate the
exponential growth in probability induced by\ the imaginary dynamic phase,
we add a factor $i$ before the Hamiltonian $H$ in the numerical simulation;
consequently $\alpha (\theta )$ becomes real and $\gamma (\theta )$ is
unaffected.

Figure. \ref{fig3} numerically and analytically exhibits the geometric phase
$e^{i\gamma (\theta )}$ on the contour $\mathcal{L}$\ with $r$ ranging from $%
0$ to $\gamma $ ($r=\gamma $ indicates that $\mathcal{L}$ is the EP ring).
The numerical results correspond with the analytical calculations in Eq (\ref%
{LI}). The oscillating frequency of the real and imaginary parts of the
geometric phase accelerates as the contour $\mathcal{L}$ approaches the EP
ring (i.e., $r=\gamma $). As a sample, the inset numerically\ presents the
fidelity when $r=\gamma /2$, which is defined as%
\begin{equation}
F(\theta )=\left\vert \left\langle \phi _{-}^{\mathrm{R}}(\theta
)\right\vert \mathcal{T}\exp [-i\int_{0}^{t}H(\theta )\mathrm{d}\theta
]\left\vert \phi _{-}^{\mathrm{R}}(0)\right\rangle \right\vert .
\end{equation}%
A fidelity of $1$ indicates that the time evolution process is adiabatic.

\section{EP surface with complicated geometry}

\label{EP surface with complicated geometry}

The distribution of the Berry curvature for the new geometry is studied in
this section. The EP surface has diverse geometries when the Hamiltonian in
Eq. (\ref{H}) is generalized to $f\left( k_{x},k_{y}\right) =m-[a\cos
k_{x}+b\cos k_{y}+c\cos \left( 2k_{x}\right) +d\cos \left( 2k_{y}\right)
+s\cos k_{z}]$\ and $s=2$. On the basis of the discussion of the physical
realization \cite{OL,LL}, the generalized Hamiltonian could be realized by
adding long-range perturbations in the $x$\ and $y$\ directions in a
periodic metallic-mesh 3D photonic crystal with $\mathcal{PT}$-symmetric
non-Hermitian elements. In Sec. \ref{Experimental scheme in
electrical circuit}, we discuss in detail how to realize an EP surface in
the electrical circuit.

Although the geometry of the EP surface changes when the parameters $\left\{
m,a,b,c,d\right\} $ vary, the EP still maintains the following features: (i)
It possesses a mirror symmetry with respect to the $k_{z}=0$\ plane. (ii)
The Hamiltonian still has two layers of EP surfaces, the two EP surfaces
touch each other when $\gamma =s$, and we discuss only the lower-layer EP
surface near $k_{z}=0$. (iii) The meridional cross-sections become irregular
circles rather than disks. (iv) The equatorial cross-section no longer has
regular geometry, but the streamlines of the Berry curvature in the
equatorial cross-section retain the feature discussed in the previous
section: they are closed curves that can be depicted according to Eq. (\ref%
{Curves}). To illustrate the distribution of the Berry curvature, the
geometries of the EP surface under two sets of parameters are studied and
the other cases share similar distributions.
\begin{figure}[th]
\includegraphics[ bb=0 0 505 616, width=8 cm, clip]{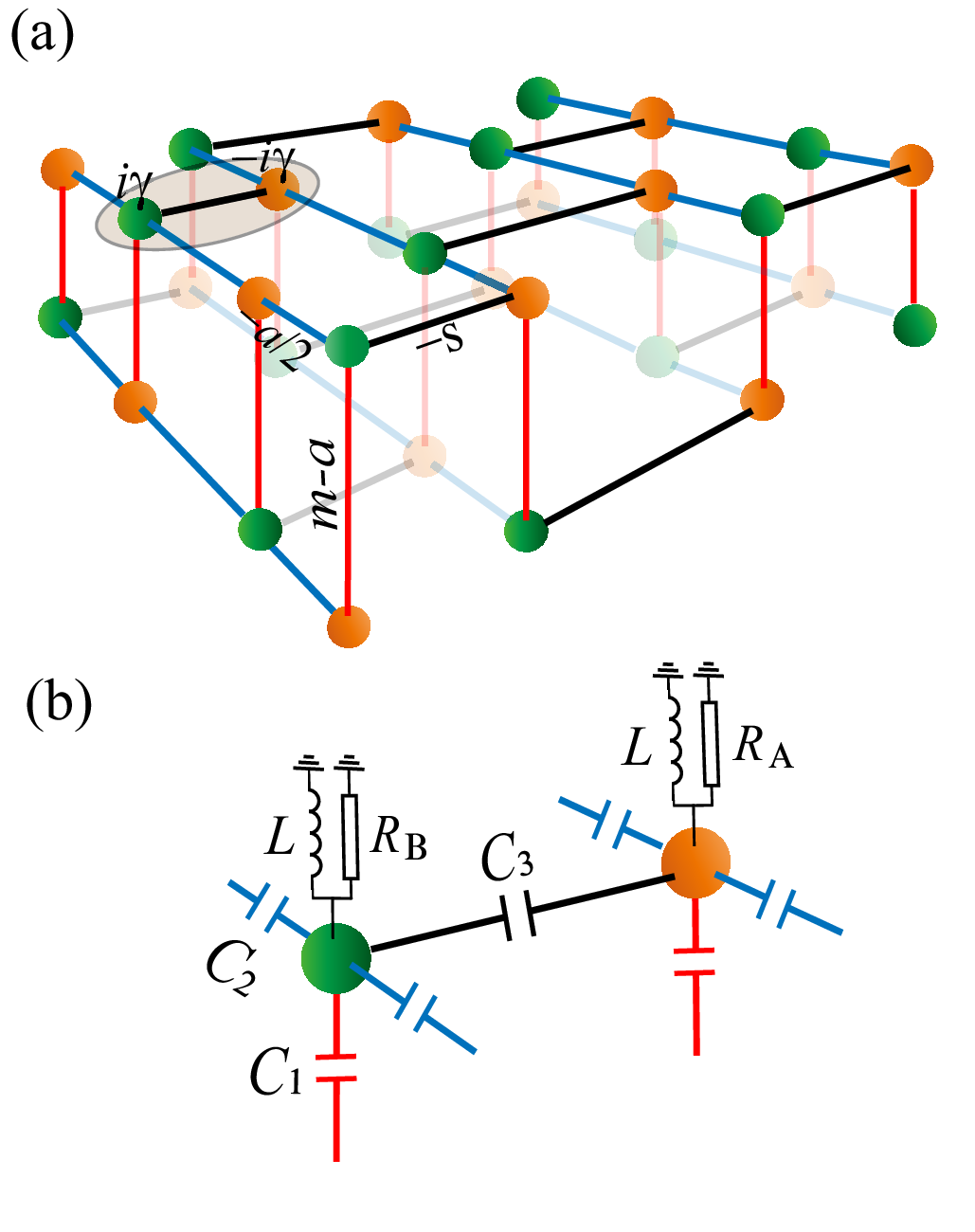}
\caption{(a) The shchematic diagram of the tight-binding lattice
in Eq. (\protect\ref{lattice}). The red, blue and black lines correspond to
the hoppings $m-a,-a/2,$ and $s$ respectively. The green and orange spheres
correspond to the gain and loss. An unit cell is marked in the shadow. (b)
The circuit elements of an unit cell. The red, blue and black capacitors,
which are denoted with $C_{1},C_{2}$ and $C_{3},$ correspond to the hoppings
in (a). The orange (green) node is connected to the ground by a inductance
and a potentiometers $R_{A}$ (negative impedance converter $R_{B}$) which
represents the gain (loss).}
\label{fig5}
\end{figure}
In the equatorial cross-section, Eq. (\ref{BerryCurvatureII}) remains valid,
and Figs. \ref{fig4}(c) and (d) depict the streamlines of the Berry
curvature. The two similar equatorial cross-sections in Fig. \ref{fig4}(c)
and (d) have five holes, and the two left or right holes are touching
(separated) in (c) [(d)]. We sort the streamlines by the number and
orientation of the holes they surround. Figure. \ref{fig4}(c) exhibits three
types of streamlines\ surrounding one hole (the center hole, and the
orientation of streamlines are anticlockwise), two holes (the two left or
right holes, clockwise), and five holes (clockwise). In Fig. \ref{fig4}(d),
in addition to the three types of streamlines, there are yet another type of
streamline surrounding one hole that flows clockwise. The appearance of the
new type of streamline is a consequence of the separation between the two
left (or right) holes in Fig. \ref{fig4}(d).

In the meridional cross-section, the Berry curvature has non-zero radial and
angular components and may be not perpendicular to the meridional
cross-section. Berry flux is nonzero if all the arrows representing the
Berry curvature point in the same direction in the meridional cross-section.
There are specific meridional cross-sections that contain no DPs or several
DPs. A natural question to ask is what the Berry curvature distribution is
in these specific meridional cross-sections. A specific meridional
cross-section containing no DPs is illustrated in Fig. \ref{fig4}(b) and the
top view of this cross-section is shown in Fig. \ref{fig4}(d) (i.e., the
yellow rectangle). In Fig. \ref{fig4}(f), the arrows indicate that the Berry
curvature points in the positive $x$-direction on the left semicircle and in
the negative $x$-direction on the right semicircle. The signs cancel out and
the Berry flux vanishes. In accordance with this distribution, inside the
yellow transparent rectangle in Fig. \ref{fig4}(d), these streamlines flow
up on the right side and down on the left, and the total flux is zero. In
Fig. \ref{fig4}(e), the non-zero winding number of the arrows indicating the
Berry connection outside the EP ring is consistent with the above conclusion
in Sec. \ref{Distribution in the meridional cross-section}, which indicates
that the winding number has no relation with the Berry flux.

\section{Experimental scheme in electrical circuit}

\label{Experimental scheme in electrical circuit}

The EP surface can be measured using an electrical circuit which
is a powerful platform for investigating topological physics \cite%
{RThomale1,RThomale2,CHLeeCP2023,CHLeePRB085426}. For the sake of convenience, this section discusses
the experimental scheme of EL in an electrical circuit, i.e., the
intersection line of the EP surface and the $S_{V}$ cross-section (see Fig. %
\ref{fig1}(b)). There are two reasons for doing this. First, the topological
properties of the EL are consistent with those of the EP surface. Second,
the experimental setup corresponding to the EL can be smoothly generalized
to that of the EP surface due to the design flexibility of the electrical
circuit.

We first show the tight-binding lattice model possessing EL. In the $S_{V}$
cross-section where $k_{x}=0$, {the auxiliary field $\mathbf{B}\left(
\mathbf{k}\right) =\left\{ B_{x}(\mathbf{k})\text{, }B_{y}(\mathbf{k}%
)\right\} $ in Eq. (\ref{H}) can be reduced to}%
\begin{equation}
\left\{
\begin{array}{l}
B_{x}=f(k_{y})-s\cos k_{z} \\
B_{y}=s\sin k_{z}%
\end{array}%
\right. ,
\end{equation}%
where $f\left( k_{y}\right) =m-a-a\cos k_{y}$, and $s=$ $1$. Substituting
the Fourier transformation
\begin{equation}
\left\{
\begin{array}{c}
a_{k_{y},k_{z}}^{\dagger }=\frac{1}{\sqrt{N}}%
\sum_{j,l}e^{ik_{y}j}e^{ik_{z}l}a_{j,l}^{\dagger } \\
b_{k_{y},k_{z}}^{\dagger }=\frac{1}{\sqrt{N}}%
\sum_{j,l}e^{ik_{y}h}e^{ik_{z}l}b_{j,l}^{\dagger }%
\end{array}%
\right.
\end{equation}%
into the core matrix $\sum_{k_{y},k_{z}}\mathbf{B}\left( k_{y},k_{z}\right)
\cdot \mathbf{\sigma }$, we get the lattice model
\begin{eqnarray}
H &=&\sum_{\mathbf{r}}((m-a)a_{\mathbf{r}}^{\dagger }b_{\mathbf{r}}-\frac{a}{%
2}\left( a_{\mathbf{r}}^{\dagger }b_{\mathbf{r+}\hat{\jmath}}+a_{\mathbf{r}%
}^{\dagger }b_{\mathbf{r-}\hat{\jmath}}\right)  \label{lattice} \\
&&-sa_{\mathbf{r}}^{\dagger }b_{\mathbf{r+}\hat{l}})+\mathrm{h.c.}+i\gamma
a_{\mathbf{r}}^{\dagger }a_{\mathbf{r}}-i\gamma b_{\mathbf{r}}^{\dagger }b_{%
\mathbf{r}}  \notag
\end{eqnarray}%
where $\mathbf{r=}x\hat{\jmath}+y\hat{l}$ is the position vector, $\hat{%
\jmath},\hat{l}$ represents the unit vectors, and the system size is $N$. A
schematic diagram of the lattice model is shown in Fig. \ref{fig5}(a){. The
hoppings }$-a/2$, $m-a$ and $-s$ are represented by the blue, red and black
lines, respectively. The on-site gains and losses are shown in orange and
green, respectively. We can extend this lattice system in the $x$-direction
to obtain a model possessing an EP surface.

The lattice system can be represented by an electrical circuit with $N$
nodes. An $N\times N$ matrix $J(\omega ,\mathbf{r}),$ termed circuit
Laplacian or admittance matrix, can be used to represent the Hamiltonian of
a tight-binding model \cite{RThomale1,RThomale2,RThomale3}. $J(\omega ,%
\mathbf{r})$ describes the voltage response $\mathbf{V}(\omega ,\mathbf{r})$
to an ac input current $\mathbf{I}(\omega ,\mathbf{r})$ according to%
\begin{equation}
\mathbf{V}(\omega ,\mathbf{r})=J(\omega ,\mathbf{r})\mathbf{I}(\omega ,%
\mathbf{r}),  \label{Laplacian}
\end{equation}%
where $\omega $ is the AC driving frequency and $r$ represents the nodes.
The vector components of $\mathbf{V}$ and $\mathbf{I}$ correspond to the
nodes or sites in the circuit. The matrix elements of $J(\omega ,\mathbf{r})$
are determined on the admittance of circuit elements between nodes or
between nodes and the ground. Figure \ref{fig5}(b) shows a schematic diagram
of the circuit elements corresponding to a unit cell.\ The lattice sites are
represented by circuit nodes. The variable hoppings, $m-a,-a/2$ and $-s$ can
be realized by tuning the capacitors $C_{\mathbf{1}},C_{\mathbf{2}}$ and $C_{%
\mathbf{3}}$, respectively. The onsite gain $i\gamma $ or loss $-i\gamma $
are realized using potentiometers or a negative impedance converter to
ground. The admittance matrix has an alternative representation in momentum
space, denoted as $J(\omega ,\mathbf{k})$. $J(\omega ,\mathbf{k})$ can be
obtained by performing $M$ linearly independent measurements in the
electrical circuit\cite{RThomale2,RThomale3,Kiessling,XZhang,SOh}, where $M$
describes the number of inequivalent nodes in the network. Each measurement
consists of a local excitation of the circuit network and a global
measurement of the voltage response, from which all the components of $%
J(\omega ,\mathbf{k})$ can be extracted. Then EL can be obtained by
diagonalizing the admittance matrix $J(\omega ,\mathbf{k}).$

\section{Discussion}

\label{Conclusion} {In summary, we have investigated the distribution of
Berry curvature inside the EP surface of $\mathcal{PT}$}-{symmetric 3D
non-Hermitian two-band systems. The EP surface acts as the separation
between the zero and non-zero Berry curvatures. }Inside a torus-like EP
surface, the distributions of Berry connections and curvatures in the
meridional and equatorial cross-sections are discussed. In the meridional
cross-section, the Berry connection serves as a planar vortex field and
diverges at the DP and EP surface. The Berry curvature has only an axial
component and diverges at the EP surface. In the equatorial cross-sections,
the Berry curvature forms the closed curves inside the EP surface. The
distributions of Berry curvature are analogous to the magnetic lines
generated by the solenoid, and the EP surface can be analogous to the
solenoids. On the basis of the distribution of the Berry curvature, we
obtain the nonquantized Berry flux. The key to identifying the zero or
non-zero Berry flux in a meridional cross-section is determining whether all
the arrows indicating Berry curvatures point in the same direction. The
numerical adiabatic evolution corresponds with the aforementioned analysis
of the nonquantized Berry flux. We also discuss the distribution of Berry
curvature in a general case in which the EP surface has more complicated
geometry. In the equatorial cross-sections, the Berry curvatures retain the
form of closed curves. The streamlines with arrows indicating the direction
of the Berry curvature are categorized by arrow orientations and the number
of the holes they surround. We discuss a scheme of realizing the EP surface
in an electrical circuit. Our findings deepen understanding of EP surfaces
and the topological properties of $\mathcal{PT}$-symmetric non-Hermitian
systems.

\acknowledgments We acknowledge the support of the National Natural Science
Foundation of China (Grants No. 12374461, No. 12222504, and No. 12305018).

\section*{Appendix}

\label{Appendix}

\subsection{Berry curvature in and out the nodal volume}

\label{Berry curvature in and out the nodal volume}

We present detailed calculations of the component for Berry connection $%
\mathcal{\vec{A}}$ and Berry curvature $\mathcal{\vec{F}}$ according to the
definitions $\mathcal{A}_{j}=\mathrm{Re}(i\left\langle \phi _{-}^{\mathrm{L}%
}\right\vert \partial _{j}\left\vert \phi _{-}^{\mathrm{R}}\right\rangle )$
and $\mathcal{F}_{j}=\partial _{l}\mathcal{A}_{i}-\partial _{i}\mathcal{A}%
_{l}$.

\textbf{(1) Outside the nodal volume} ($\gamma ^{2}<B_{x}^{2}+B_{y}^{2}$)

\textbf{a.} \textbf{Berry connection.} Substitute the Eq. (\ref{Phi_R_out})
and (\ref{Phi_L_out}) into $\mathcal{A}_{j}=\mathrm{Re}(i\left\langle \phi
_{-}^{\mathrm{L}}\right\vert \partial _{j}\left\vert \phi _{-}^{\mathrm{R}%
}\right\rangle )$, which results in%
\begin{equation}
\mathcal{A}_{j}=-\frac{1}{2}\left( \partial _{j}\alpha +\partial _{j}\beta
\right) .  \label{Aout1}
\end{equation}%
Differentiate $\tan \alpha =\gamma /\varepsilon $ where $\varepsilon =\sqrt{%
B_{x}^{2}+B_{y}^{2}-\gamma ^{2}}$%
\begin{equation}
\frac{1}{\cos ^{2}\alpha }\mathrm{d}\alpha =-\gamma \varepsilon ^{-2}\left(
\partial _{x}\varepsilon \mathrm{d}k_{x}+\partial _{y}\varepsilon \mathrm{d}%
k_{y}+\partial _{z}\varepsilon \mathrm{d}k_{z}\right) ,
\end{equation}%
move $\cos ^{2}\alpha $ to the right-hand side of "$=$" and we get%
\begin{equation}
\partial _{j}\alpha =-\frac{\left( B_{x}\partial _{j}B_{x}+B_{y}\partial
_{j}B_{y}\right) \gamma }{\varepsilon \left( B_{x}^{2}+B_{y}^{2}\right) }.
\label{alphaj}
\end{equation}%
A similar calculation performing for $\tan \beta =B_{y}/B_{x}$ yields%
\begin{equation}
\partial _{j}\beta =\frac{B_{x}\partial _{j}B_{y}-B_{y}\partial _{j}B_{x}}{%
B_{x}^{2}+B_{y}^{2}}.  \label{Beta_j}
\end{equation}%
Substitute Eq. (\ref{alphaj}) and Eq. (\ref{Beta_j}) into the Eq. (\ref%
{Aout1}), we have%
\begin{equation}
\mathcal{A}_{j}=\frac{\left( \varepsilon B_{y}+\gamma B_{x}\right) \partial
_{j}B_{x}+\left( \gamma B_{y}-\varepsilon B_{x}\right) \partial _{j}B_{y}}{%
2\varepsilon \left( B_{x}^{2}+B_{y}^{2}\right) }.  \label{Aout2}
\end{equation}

\textbf{b. Berry curvature.} Substitute Eq. (\ref{Aout1}) into $\mathcal{F}%
_{j}=\partial _{l}\mathcal{A}_{i}-\partial _{i}\mathcal{A}_{l}$,%
\begin{equation}
\mathcal{F}_{j}=\frac{1}{2}\left[ \partial _{l}\left( \partial _{i}\beta
\right) -\partial _{i}\left( \partial _{l}\beta \right) +\partial _{l}\left(
\partial _{i}\alpha \right) -\partial _{i}\left( \partial _{l}\alpha \right) %
\right] .
\end{equation}%
Firstly, we prove $\partial _{j}\left( \partial _{l}\alpha \right) -\partial
_{l}\left( \partial _{j}\alpha \right) =0$. Partial derivative of $\partial
_{j}\alpha $ is%
\begin{eqnarray}
\partial _{l}\left( \partial _{j}\alpha \right) &=&\frac{\gamma \left(
3\varepsilon ^{2}+\gamma ^{2}\right) }{\varepsilon ^{3}\left(
B_{x}^{2}+B_{y}^{2}\right) ^{2}}[B_{x}^{2}\partial _{l}B_{x}\partial
_{j}B_{x}+B_{y}^{2}\partial _{l}B_{y}\partial _{j}B_{y}  \notag \\
&&+B_{x}B_{y}\left( \partial _{l}B_{x}\partial _{j}B_{y}+\partial
_{l}B_{y}\partial _{j}B_{x}\right) ]  \notag \\
&&-\frac{\gamma }{\varepsilon \left( B_{x}^{2}+B_{y}^{2}\right) }[\partial
_{l}B_{x}\partial _{j}B_{x}+B_{x}\partial _{l}\left( \partial
_{j}B_{x}\right)  \label{DD_alpha} \\
&&+B_{y}\partial _{l}\left( \partial _{j}B_{y}\right) +\partial
_{l}B_{y}\partial _{j}B_{y}].  \notag
\end{eqnarray}%
$\partial _{j}\left( \partial _{l}\alpha \right) $ can be obtained by
swapping $j$ with $l$ in Eq. (\ref{DD_alpha}) and $\partial _{l}\left(
\partial _{j}\alpha \right) $ has the same expression with $\partial
_{j}\left( \partial _{l}\alpha \right) $, which means%
\begin{equation}
\partial _{l}\left( \partial _{j}\alpha \right) -\partial _{j}\left(
\partial _{l}\alpha \right) =0.
\end{equation}%
Secondly, we prove that $\partial _{l}\left( \partial _{j}\beta \right)
-\partial _{j}\left( \partial _{l}\beta \right) =0$. Partial derivative of $%
\partial _{j}\beta $ is%
\begin{eqnarray}
\partial _{l}\left( \partial _{j}\beta \right) &=&\frac{1}{\left(
B_{x}^{2}+B_{y}^{2}\right) ^{2}}[2B_{y}B_{x}\left( \partial
_{l}B_{x}\partial _{j}B_{x}-\partial _{l}B_{y}\partial _{j}B_{y}\right)
\notag \\
&&+\left( B_{y}^{2}-B_{x}^{2}\right) \left( \partial _{l}B_{x}\partial
_{j}B_{y}+\partial _{l}B_{y}\partial _{j}B_{x}\right) \\
&&+\left( B_{x}^{2}+B_{y}^{2}\right) \left( B_{x}\partial _{l}(\partial
_{j}B_{y})-B_{y}\partial _{l}(\partial _{j}B_{x}\right) )].  \notag
\end{eqnarray}%
\qquad $\partial _{j}\left( \partial _{l}\beta \right) $ can be obtained by
swapping $j$ with $l$ and it is not difficult to check that $\partial
_{j}\left( \partial _{l}\beta \right) =\partial _{l}\left( \partial
_{j}\beta \right) $, which is%
\begin{equation}
\partial _{l}\left( \partial _{j}\beta \right) -\partial _{j}\left( \partial
_{l}\beta \right) =0.  \label{DDBeta}
\end{equation}%
So we prove that $\mathcal{F}_{j}=0$, i.e. $\mathcal{\vec{F}}=\nabla \times
\mathcal{\vec{A}}=0.$

\textbf{(2) Inside the nodal volume} ($\gamma ^{2}>B_{x}^{2}+B_{y}^{2}$)

\textbf{a. Berry connection.} Substituting Eq. (\ref{Phi_R_in}) and (\ref%
{Phi_L_in}) into $\mathcal{A}_{j}=\mathrm{Re}(i\left\langle \phi _{-}^{%
\mathrm{L}}\right\vert \partial _{j}\left\vert \phi _{-}^{\mathrm{R}%
}\right\rangle )$, the component of $\mathcal{\vec{A}}$ is%
\begin{eqnarray}
\mathcal{A}_{j} &=&-\mathrm{Re}[\frac{1}{\Omega }\left( \eta ^{2}\partial
_{j}\beta +i\eta \partial _{j}\eta \right) ]  \label{Im} \\
&=&-\frac{\eta ^{2}}{\Omega }\partial _{j}\beta  \notag \\
&=&\frac{\left( \varepsilon -\gamma \right) \left( B_{x}\partial
_{j}B_{y}-B_{y}\partial _{j}B_{x}\right) }{2\varepsilon \left(
B_{x}^{2}+B_{y}^{2}\right) }  \label{BerryCon}
\end{eqnarray}%
where $\varepsilon =\sqrt{\gamma ^{2}-B_{x}^{2}-B_{y}^{2}},\eta =(\gamma
-\varepsilon )/\sqrt{B_{x}^{2}+B_{y}^{2}}$, and $\Omega =2\eta \varepsilon /%
\sqrt{B_{x}^{2}+B_{y}^{2}}.$

\textbf{b. Berry curvature.} Differentiate $\mathcal{A}_{j}$ to obtain%
\begin{equation}
\partial _{l}\mathcal{A}_{j}=\frac{\varepsilon -\gamma }{2\varepsilon }%
\partial _{l}(\partial _{j}\beta )-\partial _{l}(\frac{\gamma }{2\varepsilon
})\left( \partial _{j}\beta \right) .
\end{equation}%
$\partial _{j}\mathcal{A}_{l}$ can be obtained by swapping the indexes $%
l\leftrightarrow j,$ $\partial _{j}\mathcal{A}_{l}=[\partial _{j}(\partial
_{l}\beta )](\varepsilon -\gamma )/(2\varepsilon )-\partial _{j}(\varepsilon
^{-1})\left( \gamma \partial _{l}\beta /2\right) $. Therefore the component
of berry curvature can be calculated%
\begin{equation}
\mathcal{F}_{i}=\partial _{l}\mathcal{A}_{j}-\partial _{j}\mathcal{A}_{l}=%
\frac{\gamma }{2\varepsilon ^{2}}\left( \partial _{l}\varepsilon \partial
_{j}\beta -\partial _{j}\varepsilon \partial _{l}\beta \right) ,
\end{equation}%
and Eq. (\ref{DDBeta}) is used in the calculation. Then move $\partial
_{j}\varepsilon =-\varepsilon _{k}^{-1}\left( B_{x}\partial
_{j}B_{x}+B_{y}\partial _{j}B_{y}\right) $ and Eq. (\ref{Beta_j}) into
aforementioned equation, we have%
\begin{equation}
\partial _{l}\mathcal{A}_{j}-\partial _{j}\mathcal{A}_{l}=\frac{\gamma }{%
2\varepsilon ^{3}}\left( \partial _{l}B_{y}\partial _{j}B_{x}-\partial
_{l}B_{x}\partial _{j}B_{y}\right) ,
\end{equation}%
that is%
\begin{equation}
\mathcal{F}_{j}=\partial _{l}\mathcal{A}_{i}-\partial _{i}\mathcal{A}_{l}=%
\frac{\gamma }{2\varepsilon ^{3}}\left( \partial _{l}B_{y}\partial
_{i}B_{x}-\partial _{l}B_{x}\partial _{i}B_{y}\right) .  \label{BerryCur}
\end{equation}

\subsection{Berry connection and curvature in the meridional cross-section}

\label{Berry connection and curvature in the meridional cross-section}

This section gives the detials of calculating Eq. (\ref{BCpolar}) and Eq. (%
\ref{Bcurvature}) under the constraint of condition $0<r<\gamma .$

The EP surface is in the form of a torus under the condition $m=\sqrt{\left(
2-\gamma \right) /\gamma }+1$, $a=(m-1+\sqrt{\left( m-1\right) ^{2}-\gamma
^{2}+4})/\left( 4-\gamma ^{2}\right) $, as presented in \ref{fig1} (a). The
schematic of the meridional cross-section is presented in Fig. \ref{fig1}
(b). The center of the circle $\mathcal{O}$($k_{y_{0}},$ $0$) is the DP for
the Hermitian case; therefore, $k_{y_{0}}$ meets the condition $%
B_{x}(0,y_{0},0)=0$, which yields%
\begin{equation}
a\cos k_{y_{0}}=m-a-1,  \label{IV_1}
\end{equation}%
and%
\begin{equation}
a^{2}\sin ^{2}k_{y_{0}}=\left( m-1\right) \left( 2a-m+1\right) .
\label{IV_2}
\end{equation}%
Replace $m$ and $a$ with $\gamma $ in Eq. (\ref{IV_2}), we have $a\sin
k_{y_{0}}=\pm 1$. Here we discuss the case of $a\sin k_{y_{0}}=-1$. In polar
coordinates, $\left( k_{y},k_{z}\right) =\left( k_{y_{0}}+r\cos \theta
,r\sin \theta \right) $, where $r$ is small. Using the above parameter
settings and taking the Taylor expansion, the Hamiltonian in Eq. (\ref%
{Hamiltonian}) can be rewritten as%
\begin{equation}
H=\left(
\begin{array}{cc}
i\gamma & re^{i\theta } \\
re^{-i\theta } & -i\gamma%
\end{array}%
\right)
\end{equation}%
and the Berry connection in Eq. (\ref{AcompIn}) and Berry curvature in Eq. (%
\ref{BcurIn}) can be reduced into%
\begin{equation}
\left\{
\begin{array}{c}
\mathcal{A}_{x}=0 \\
\mathcal{A}_{y}\approx \frac{\varepsilon -\gamma }{2\varepsilon r}\sin \theta
\\
\mathcal{A}_{z}\approx -\frac{\varepsilon -\gamma }{2\varepsilon r}\cos
\theta%
\end{array}%
\right. ,\text{ and }\left\{
\begin{array}{c}
\mathcal{F}_{x}\approx \frac{\gamma }{2\varepsilon ^{3}} \\
\mathcal{F}_{y}=\mathcal{F}_{z}=0%
\end{array}%
\right.
\end{equation}%
By using a coordinate transformation of%
\begin{equation}
\left(
\begin{array}{c}
\mathcal{A}_{r} \\
\mathcal{A}_{\mathcal{\theta }}%
\end{array}%
\right) =S\left(
\begin{array}{c}
\mathcal{A}_{y} \\
\mathcal{A}_{z}%
\end{array}%
\right) ,
\end{equation}%
where
\begin{equation}
S=\left(
\begin{array}{cc}
\cos \mathcal{\theta } & \sin \mathcal{\theta } \\
-\sin \mathcal{\theta } & \cos \mathcal{\theta }%
\end{array}%
\right) ,  \label{Smatrix}
\end{equation}%
the form for Berry connection and curvature in polar coordinates,%
\begin{equation}
\left\{
\begin{array}{c}
\mathcal{A}_{r}=\mathcal{A}_{x}=0 \\
\mathcal{A}_{\mathcal{\theta }}=\frac{\sqrt{\gamma ^{2}-r^{2}}-\gamma }{%
2\varepsilon r}%
\end{array}%
\right. \text{ and }\left\{
\begin{array}{c}
\mathcal{F}_{r}=\mathcal{F}_{\mathcal{\theta }}=0 \\
\mathcal{F}_{x}=\frac{\gamma }{2\varepsilon ^{3}}%
\end{array}%
\right. .
\end{equation}%
can be easily obtained.

\section{Berry connection and Berry curvature defined under the Dirac norm}

This section gives the expressions for the Berry connection and Berry
curvature under the Dirac orthonormal basis in the broken region. As the
purpose of this section is to compare the results with those obtained under
the definition of biorthogonal bases sets, and the details of calculation is
similar to that in Sec. \ref{Berry curvature in and out the nodal volume},
therefore only the results will be presented.

The expression for the eigenstates of the Hamiltonian in Eq. (\ref%
{Hamiltonian}) is%
\begin{equation}
\left\vert \phi _{+}^{\mathrm{R}}\right\rangle =\frac{1}{\sqrt{\Lambda }}%
\left(
\begin{array}{c}
\eta e^{i\left( \frac{\pi }{2}-\beta \right) } \\
1%
\end{array}%
\right)
\end{equation}%
corresponding to the eigenvalue $-i\sqrt{\gamma ^{2}-B_{x}^{2}-B_{y}^{2}},$%
where%
\begin{equation}
\Lambda =\eta ^{2}+1,\eta =\frac{\gamma +\sqrt{\gamma
^{2}-B_{x}^{2}-B_{y}^{2}}}{\sqrt{B_{x}^{2}+B_{y}^{2}}}.
\end{equation}%
Berry connection can be defined by%
\begin{equation}
\mathcal{\vec{A}}_{\vec{k}}^{\mathrm{d}}=i\left\langle \phi _{+}^{\mathrm{R}%
}\right\vert \nabla _{\vec{k}}\left\vert \phi _{+}^{\mathrm{R}}\right\rangle
=\mathcal{A}_{k_{x}}^{\mathrm{d}}\mathbf{e}_{\mathbf{x}}+\mathcal{A}%
_{k_{y}}^{\mathrm{d}}\mathbf{e}_{\mathbf{y}}+\mathcal{A}_{k_{z}}^{\mathrm{d}}%
\mathbf{e}_{\mathbf{y}},
\end{equation}%
where the component reads%
\begin{equation}
\mathcal{A}_{j}^{\mathrm{d}}=\frac{\varepsilon +\gamma }{2\gamma \left(
B_{x}^{2}+B_{y}^{2}\right) }(B_{x}\partial _{j}B_{y}-B_{y}\partial _{j}B_{x})
\end{equation}%
and Berry curvature can be defined by%
\begin{equation}
\mathcal{F}_{j}^{\mathrm{d}}=\partial _{l}\mathcal{A}_{i}^{\mathrm{d}%
}-\partial _{i}\mathcal{A}_{l}^{\mathrm{d}}=\frac{\partial _{i}B_{y}\partial
_{l}B_{x}-\partial _{l}B_{y}\partial _{i}B_{x}}{2\gamma \varepsilon }.
\end{equation}%
The denominator of $\mathcal{A}_{j}^{\mathrm{d}}$ has one less factor of $%
\varepsilon $ compared to the denominator of $\mathcal{A}_{j}$ in Eq. (\ref%
{AcompIn}), which leads a convergent $\mathcal{A}_{j}^{\mathrm{d}}$ at the
EP surface.

\end{document}